\documentclass[10pt,a4paper]{article}
\usepackage[a4paper]{geometry}
\usepackage[T1]{fontenc}
\usepackage{mathpazo}
\usepackage[latin1]{inputenc}							
\usepackage{amsmath}									
\usepackage{xcolor,graphicx}
\definecolor{bl}{rgb}{0.0,0.2,0.6}
\definecolor{nicered}{rgb}{.647,.129,.149}

\usepackage{caption}
\DeclareCaptionFont{myblue}{\color{bl}}
\usepackage{hyperref}
\hypersetup{
    bookmarks=true,         
    pdftoolbar=true,        
    pdfmenubar=true,        
    pdffitwindow=false,     
    pdfstartview={FitH},    
    pdfauthor={Thomas Bury},     
    pdfnewwindow=true,      
    colorlinks=true,       
    linkcolor=bl,          
    citecolor=blue ,        
    filecolor=red!90,      
    urlcolor=blue ,          
}

\usepackage{sectsty}
\usepackage[compact]{titlesec}
\titleformat{\section}{\color{nicered}\large\bf}{\thesection}{1em}{}
\titleformat{\subsection}{\color{nicered}\normalsize\bf}{\thesubsection}{1em}{}
\titleformat{\subsubsection}{\color{nicered}\normalsize\bf}{\thesubsubsection}{1em}{}

\makeatletter							
\def\printtitle{
    {\color{bl} \flushleft \huge \@title\par}}		
\makeatother							

\title{Predicting trend reversals using market instantaneous state}

\makeatletter							
\def\printauthor{
    {\hfill\parbox[b]{0.90\textwidth}{\flushleft \small \@author}}}				
\makeatother							

\author{%
	\textbf{\large Thomas Bury} \\[1\baselineskip]
    Service OPERA (CP194/5), Universit\'e libre de Bruxelles,\\
    Avenue F.D. Roosevelt 50, 1050 Brussels, Belgium\\
	Email:tbury@ulb.ac.be \\
	}

\usepackage{fancyhdr}
\usepackage{lastpage}	
\usepackage[runin]{abstract}		        
\abslabeldelim{:\quad}						%
\begin{document}
\printtitle

\printauthor

\begin{abstract}Collective behaviours taking place in financial markets reveal strongly correlated states especially during a crisis period. A natural hypothesis is that trend reversals are also driven by mutual influences between the different stock exchanges. Using a maximum entropy approach, we find coordinated behaviour during trend reversals dominated by the pairwise component. In particular, these events are predicted with high significant accuracy by the ensemble's instantaneous state.
\end{abstract}

\hrule
\footnotesize
\tableofcontents
\vspace{1em}
\hrule
\vspace{1em}
\normalsize
\section{Introduction}
\label{intro}
Despite abundant research focusing on estimating the level of stock returns, there are few studies examining the predictability of the sign of financial asset movements even though evidence of predictability of direction of excess return exists (the difference between returns and a defined benchmark) \cite{Chris,Moat,PreisGoogle,PreisComplex}. The herd behaviours of traders may explain this partial predictability \cite{Lux,Contherd,SornBook,Feng}. The orientation is an interesting quantity for capital allocation between different financial products but also because it allows the study of  collective behaviours as in neural networks and magnetic materials \cite{Fischer,Schneidman,moi2}.

Existing approaches of trend prediction are based on the connection between return volatility, skewness, kurtosis and return sign \cite{Chris}. Autologistic models (logistic models including past returns in a binary model) \cite{An} and a decomposition of the trade-to-trade price increments into three components (activity, direction and size) were considered as well as probit models with various commonly used financial variables as explanatory variables \cite{ny}. The problem with these models may be the use of a particular data generation process or the identification of relevant financial variables in the regression model. Moreover, observed collective behaviours in financial markets highlight the requirement of a multivariate approach to capture co-movements that are a key feature to explain synchronization, order, non-random correlations and predictability \cite{Dal,Laloux,OnnelaPRE,mol,Plerou99}. We believe that any model intended to predict a financial quantity, like the sign of stock returns, should therefore be multivariate. Here we propose a statistical data-based model capturing almost all the correlation structure of a financial market, the so-called pairwise maximum entropy model \cite{moi1}. This qualitative model does not rely on a particular data generation dynamics, uses only a data-driven approach based on internal inputs (present and past returns) and takes into account co-movement.

The use of pairwise maximum entropy (maxent) models has led to a fruitful description of complex systems, particularly in phase transition and magnetic materials (Ising models and spin glasses) \cite{Fischer,Stanley}, but also in neuroscience \cite{Schneidman}.  They are related to graphical models, Boltzmann machines, error correcting codes, logistic regression, etc. \cite{Opper}. Maxent models are much more than models recovering moments from data, they are powerful effective models describing collective behaviors. However, one must pay attention to the scaling of parameters capturing co-movements (pairwise influences). In real neural networks, they seem to be size independent. Increasing the size is equivalent to lowering the temperature and freezing is prevented by the presence of negative pairwise couplings \cite{Schneidman} whereas in financial networks, couplings seem to scale as the inverse of the network size leading to a mean-field description \cite{moi2}.

The aim of the maxent approach is two-fold: use a statistical framework avoiding as much as possible any assumption and study the importance of co-movement (necessity of a multivariate approach), especially in spatial predicting stock market orientation. We found that instantaneous conditional transitions (\emph{spatial} predictions) are able to predict in average $83\%$ of market place reversals which is far better than the individual model, thereby showing the importance of co-movements. Accuracy drops to $73\%$ for the components of the Dow Jones index. Such deviation may be induced by the lower correlations and the lack of large enough samples. Furthermore, we showed that history does not seem to improve the accuracy either by a genuine lack of memory or by a finite size effect in the parameter inference. These results suggest that some collective dynamics drives the global market trend \cite{Laloux,Plerou99}. They constitute another evidence of coordinated behaviours in financial markets. Moreover, they show that these collective modes are partially responsible for predictability of stock market orientation \cite{Mantegna,mol}.

We note that if a \emph{good} approximation of the collective dynamics was known together with dependencies between economic quantities, it would certainly lead to better predictions than those obtained by this simple autologistic model as it is the case in the related field of neural networks \cite{pill} and in econometric approaches \cite{An,ny}. We propose that this model serves as a benchmark with which to compare results of more sophisticated models embedding a real economic description.

\section{Collective states}
We consider a set of 8 major European indices of the Eurozone (AEX, BEL, CAC, DAX, EUROSTOXX, FTSE, IBEX, MIB) observed during a ten year long daily time series including two large crises (2008 subprimes and Euro-debt crises). The data were cleaned up to ensure simultaneity of the different time series (see appendix). An orientation reversal (or a \emph{flip}) is a trend reversal in two consecutive observed trading days. More precisely we consider daily returns (without the overnight period) defined as $r_{i,t}=(p^{\mathrm{c}}_{i,t}-p^{\mathrm{o}}_{i,t})/p^{\mathrm{o}}_{i,t}$, where $p^{\mathrm{c}}_{i,t}$ is the closing price of the $i$th stock of the  period $t$ and $p^{\mathrm{o}}_{i,t}$ the opening one. The index $i=1,\ldots,N$ labels assets ($N$ is the total number of assets). The index $t=1,\ldots,T$ labels time periods ($T$ is the total number observed periods). They can be rewritten as $r_{i,t}=s_{i,t}|r_{i,t}|$ where the binary variable $s_{i,t}\in\{-1,1\}$ is the sign or orientation of the index $i$ at period $t$. An orientation change occurs if $s_{i,t+1}=-s_{i,t}$. Such reversals are expressed as a binary variable $\mathbf{1}_{[s_{i,t+1}=-s_{i,t}]}$.  We consider the binary part of returns, $1$ for a positive return and $-1$ for a negative one. The resulting time series are strongly correlated, off-diagonal correlation coefficients lie between $0.43$ and $0.74$. We consider market orientation reversal as a multivariate stochastic process. This process can be decomposed in two main components, the instantaneous (influence within the defined time-bin unit or \emph{spatial} dependence) and the causal (temporal dependence) statistical dependencies among different market places.
The study of collective state and conditional flipping probability, causal and instantaneous, requires estimation of the probability distribution of a potentially high-dimensional system ($\sim N^{2}$ parameters) which is in general intractable without further constraints.
A way to tackle this problem is to use the maximum entropy principle \cite{Jaynes,Cover} restricted to second-order moments to infer a statistical model. One obtains a multivariate autologistic model (or Ising-like model). For our data, the pairwise statistical dependencies account for $95\%$ of all statistical dependencies as measured by the multi-information criterion \cite{Schneid_Multi,moi1} and this model is suitable for the description of collective behaviors. The resulting pairwise distribution is given by

\begin{equation}
p_{2}(s_{1,t};\cdots;s_{N,t})=\mathcal{Z}^{-1}\exp\left(\frac{1}{2}\sum_{i, j=1}^{N}J_{ij}s_{i,t}s_{j,t}+\sum_{i=1}^{N}h_{i}s_{i,t}\right)\label{Lagrange}
\end{equation}
where the binary variables $s_{i}\in \{-1,1\}$ describe the orientation of market places (respectively bearish or bullish), $\mathcal{Z}$ is a normalizing constant. The parameters $\{h_{i}\}$ and $\{J_{ij}\}$ are respectively Lagrange multipliers associated with first and second order constraints.
In this framework the instantaneous dependencies among indices (or stocks) are given in terms of conditional flipping probabilities of a given index. The flipping rate is given by

\begin{equation}\label{InstPr}
p(-s_{i,t-1}=s_{i,t}|\mathbf{s}_{-i,t})=\frac{\exp\left(-s_{i,t-1}\sum_{ j\neq i}J_{ij}s_{j,t}-h_{i}s_{i,t-1}\right)}{\exp\left(-\sum_{ j\neq i}J_{ij}s_{j,t}-h_{i}\right)+\exp\left(\sum_{j\neq i}J_{ij}s_{j,t}+h_{i}\right)}
\end{equation}
where $\mathbf{s}_{-i,t}$ is the observed market configuration at period $t$, excluding the $i$th entity.

\begin{figure}
\resizebox{1\columnwidth}{!}{%
  \includegraphics{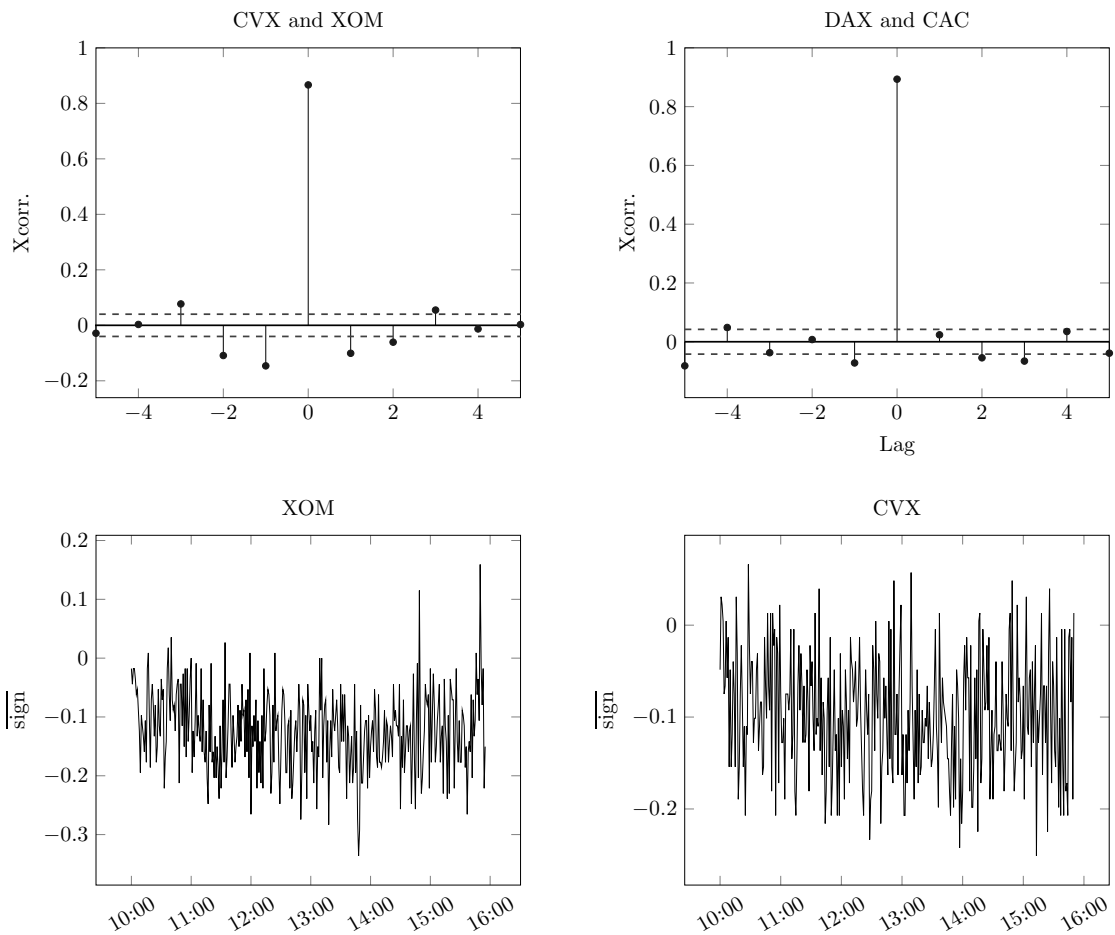}
}
\caption{\textbf{Top}: cross-correlogram between orientation of CVX and XOM (left) and between CAC and DAX indices. CVX and XOM are two main oil companies, 2500 daily returns have been used.
\textbf{Bottom}: the sign as a function of time for intraday data at minute sampling for XOM (left) and CVX (right). The bar stands for the temporal mean over 225 trading days (between March 2011 and May 2012).}
\label{fig:XCorr}     
\end{figure}

One can enquire whether considering past states could help to predict flipping events. The conditional flipping probability (\ref{InstPr}) can be modified to include some memory and is given by

\begin{equation}\label{HistPr}
p(-s_{i,t-1}=s_{i,t}|\mathcal{H}_{t}^{T})=\frac{1}{2}
\left[1-s_{i,t-1}\tanh\left(\sum_{ j\neq i}J_{ij}s_{j,t}+h_{i}
+\sum_{\tau=1}^{T}\sum_{j}K_{ij}^{\tau}s_{j,t-\tau}\right)\right]
\end{equation}

where the history $\mathcal{H}_{t}^{T}$ denotes the sequence $(\mathbf{s}_{-i,t}; \mathbf{s}_{t-1};\ldots; \mathbf{s}_{t-T})$. We expect minor difference with the memoryless case since sign autocorrelations and pairwise cross-correlations are known to be insignificant for any lag (except the first one in some case) \cite{Cont,BouchaudBook} at the contrary of their absolute values \cite{Pod}; cross-correlations between CAC and DAX indices and between CVX and XOM stocks are illustrated in Fig-\ref{fig:XCorr}, for instance. However cross-correlations measure linear or monotonic dependencies. More sophisticated statistical relationships may exist. Maxent models are supposed to capture them as the entropy and related quantities provide a more general way to capture statistical dependencies \cite{Cover}.
Furthermore, we can check if our model is able to \emph{forecast} sign of returns by checking if the predictive power is significantly larger than $50\%$ when we consider only \emph{past} information (and so, make profit). We will see that it is not the case. This result is in line with the weak efficient market hypothesis (roughly: one cannot forecast the \emph{sign} of excess returns using only past returns)\cite{Fama}.
In the following, we restrict ourself to two time-lags (since more lags mean more parameters to estimate and decrease the prediction power in our data set). For higher sampling frequency (here, the minute timescale), specific features may influence the results. Firstly, prices move discretely (jumps) as they can only vary by $1$ cent increment. We have not considered this issue in the analysis but we considered highly capitalized and very liquid assets which can limit the impact of the so-called market structure noise. Secondly, the absolute intraday returns draw a concave curve with a minimum reached at lunch time (intraday seasonality). This deterministic pattern is observed throughout the market \cite{Andersen}. We looked for such seasonality in the sign of return. The temporal mean over $225$ trading days of the intraday signs (between 10:00 am and 4:00 pm) is illustrated in bottom panels of Fig-\ref{fig:XCorr}. There is not a clear deterministic pattern neither in the time domain nor in the frequency domain (not illustrated here), meaning there is not a preferential direction of trades (sell or buy) at the opening and closing of a trading day.

Lagrange parameters were estimated by a regularized pseudo-maximum likelihood method (rPML) (see appendix) \cite{Aurell}. Once they were estimated, the flipping probability is obtained using (\ref{InstPr}) or (\ref{HistPr}).

However the distinction between statistical dependencies induced by \emph{correlated} common inputs $\{h_{i}\}$ and genuine pairwise ones should be done. In the pairwise maxent framework, if an input (says $h_{j}$) is dependent of another one (says $h_{i}$) this can lead to a non-diagonal covariance even if $J_{ij}$ are set to zero.

\section{Results}

\subsection{Indices set}
\begin{figure}
\resizebox{1\columnwidth}{!}{%
  \includegraphics{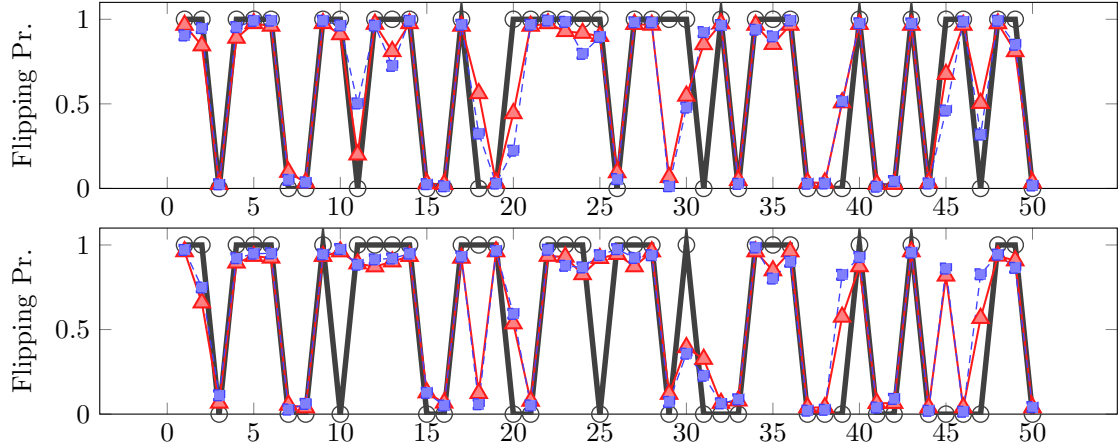}
}
\caption{Predicted series for the CAC (top) and DAX (bottom) indices. The black circles represent the actual flipping time series for 50 out-of-samples trading days. The red full line (triangles) illustrates the memoryless flipping probability and the blue dashed line (square points) the flipping probability including two time-lags.}
\label{fig:InstPr}
\end{figure}

First of all, we perform a preliminary test. We infer Lagrange parameters on a large time-window (more than 2000 trading days) and we computed flipping probabilities for 50 out-of-sample consecutive trading days using either instantaneous empirical data $\textbf{s}_{-i,t}$ in (\ref{InstPr}) or empirical sequence $\mathcal{H}_{t}^{T}$ in (\ref{HistPr}). The results for CAC and DAX indices are illustrated in Fig-\ref{fig:InstPr}.

\begin{figure}
\begin{center}
\resizebox{.75\columnwidth}{!}{%
  \includegraphics{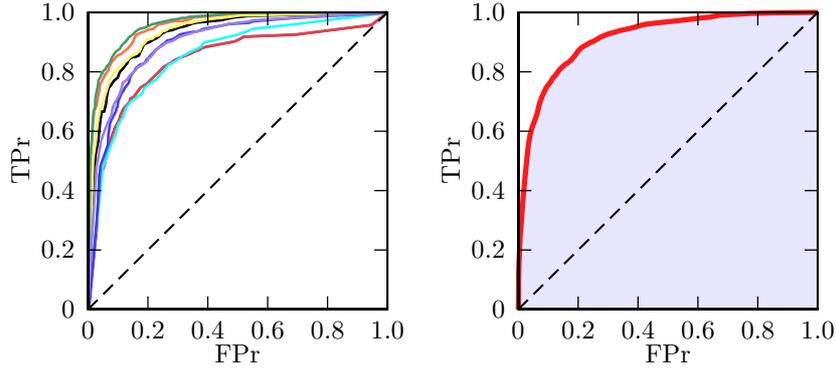}
}
\caption{Prediction of a single market place trend reversal. The receiver operating characteristics (ROC) curves for 8 indices (left) and the resulting mean ROC curve (right). The ROC curve illustrates the true positive rate (TPr) as a function of the false positive rate (FPr). These curves were obtained with a ten-fold cross-validation scheme on the set of the 8 European indices. The shaded area below the mean ROC curve illustrates the \emph{area under the curve} (AUC).}
\label{fig:ROC}
\end{center}
\end{figure}

Both autologistic models give similar results close to the actual time series.
To assess the efficiency of instantaneous and historical models, we compare the true-positive (predicting a flip which actually occurs) rate to the false-positive (predicting a flip which does not occur) rate. Ideally, a good classifier is supposed to have a large accuracy, but also a large true-positive rate together with a low false-positive rate.
To evaluate these quantities, we consider the confusion matrix for varying detection level. The detection level $\alpha$ is the threshold value such that the flipping is considered as a true event if flipping probability is larger than $\alpha$. We used the so-called ROC (receiver operating characteristics) curves to illustrate the predictive power of the classifier \cite{fawcett}. We used a ten-fold cross-validation scheme to compare the performance of both methods on out-of-sample events because the fitting may lead to accurate predictions if predicted states are in the training set (in-sample) but poor predictions on the validation set (out-of-sample). The sample is divided in learning and testing blocks. Parameters are estimated on $90\%$ of the total amount of data (learning block). The prediction is performed on the validation sample ($10\%$ of the data set) using empirical orientations $\textbf{s}_{-i,t}$ (or $\mathcal{H}_{t}^{T}$) belonging to the testing block to infer $s_{i,t}$. The true-positive, false-positive and accuracy rates are measured for each validation fold and are averaged over the ten folds. The ROC curves are illustrated in Fig-\ref{fig:ROC}.

\begin{figure}
\begin{center}
\begin{center}
\resizebox{0.60\columnwidth}{!}{%
  \includegraphics{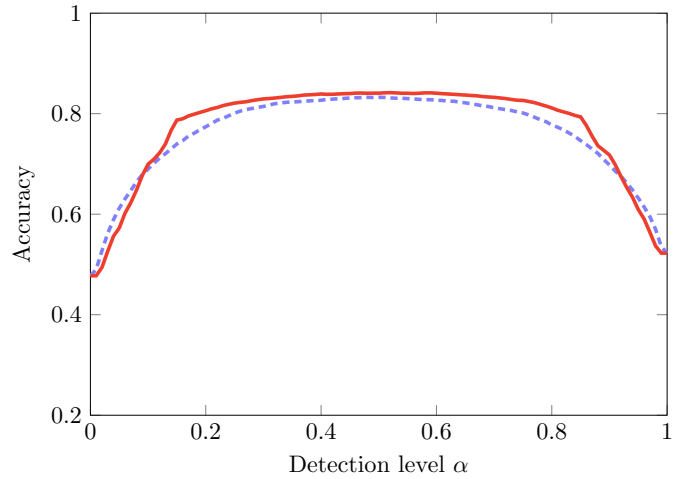}
}
\end{center}
\caption{The mean accuracy as a function of the detection level for the set of 8 European indices. The accuracy of the memoryless case is illustrated by the full line and the causal model by the dashed line.}
\label{fig:Acc}
\end{center}
\end{figure}

For the memoryless model (\ref{InstPr}), the mean true-positive rate is about $76\%$ for less than $10\%$ false-positive rate. Another summary quantity is the \emph{area under the curve} (AUC). The random guessing produces the diagonal line and thus an AUC$=0.5$. A good classifier should have an AUC close to 1. The AUC may be interpreted as the probability that the model will assign a larger flipping probability to a randomly chosen sample containing a positive event. The AUC, illustrated by the shaded area in Fig-\ref{fig:ROC}, is equal to $0.914\pm0.042$ (mean $\pm$ s.d.). The lowest AUC for the set of 8 indices is equal to $0.849$ and the largest to $0.960$. We consider also the accuracy of the prediction as a function of the chosen detection level. The accuracy is  the number of true predictions divided by the total number of events. The mean accuracy versus the detection level is illustrated in Fig-\ref{fig:Acc}. The maximum mean accuracy is equal to $83\%$. In average $83\%$ of the total number of events were correctly predicted. The lowest value of these maximal rates is equal to $78\%$ and the largest maximal rate to $89\%$.

For the historical model (\ref{HistPr}), the mean true-positive rate is about $75\%$ for less than $10\%$ false-positive rate and the resulting AUC is $0.902\pm0.050$. This mean value is not included in the $96\%$ confidence interval of the memoryless AUC but the relative deviation between both AUC mean values is only $1.3\%$. The lowest AUC for the set of 8 indices is equal to $0.849$ and the largest to $0.960$ as for the memoryless model. The maximum mean accuracy is equal to $83\%$. In average $83\%$ of the total number of events were correctly predicted. The lowest value of these maximal rates is equal to $78\%$ and the largest maximal rate to $89\%$. In the appendix, we discuss the effect of the noise in parameters estimation and the comparison of the empirical accuracy to the accuracy of artificial data generated by a truly memoryless pairwise model. The empirical accuracy is close to the largest achievable value given finite sample and system sizes.

We note that the independent instantaneous model gives a very poor result and is nearly a random guessing (AUC $=0.51$). The independent model is defined by setting $\textbf{J}$ to zero in the instantaneous model. If we consider the historical model (\ref{HistPr}) without the instantaneous part, the maximal average accuracy is only $53\%$. Therefore, we conclude that the most important component in the pairwise maxent model is the one capturing instantaneous co-movements (here, the intraday co-movements). We note that the econometrical model detailed in \cite{ny} correctly forecasts $59\%$ of out-of-sample events showing the importance of the knowledge, even partial, of the fundamental relationships between economic quantities.
Last, we note that a drawback of the historical model is the multiplication of parameters to be estimated. Each added time-step brings $(N^2-N)/2$ more parameters for each matrix $\mathbf{K}^{\tau}$. However the sum should be truncated at an optimal lag (the one where the accuracy reaches its maximum value, for instance).
We conclude that the most significant part of the prediction model is the one capturing instantaneous co-movements, in line with the efficient market hypothesis.

\subsection{Dow Jones}

The Dow Jones is an index regrouping highly capitalized US companies (AA, AXP, BA, BAC, CAT, CSCO, CVX,	DD,	DIS, GE, HD, HPQ, IBM, INTC, JNJ,	KFT, KO, MCD, MMM, MRK,	MSFT, PFE, PG, T, TRV, UTX,	VZ,	WMT, XOM).
We consider two different timescales: daily and 1 minute price sampling rates. The sample size for the daily sampling is about $2500$ trading days and $3\times10^4$ points for the minute timescale.
In this application, there are two main issues. For a satisfactory parameters estimation, we need large samples. A direct sampling would require a sample length several times larger than the total number of configurations $2^N$, which is huge for the Dow Jones ($\sim 10^9$ points which means 5 thousand trading years at this timescale). For the rPML method, the reconstruction may be done with fewer points, but still with large sample lengths $10^6$ to $10^8$ for a system size $N=64$ \cite{Aurell}. Secondly, the typical correlation coefficients between orientations are smaller than those of market places. The issues are thus twofold: parameter estimation may be flawed and low correlations may lead to intrinsically lower predictive power than in indices set analysis.


\begin{figure}
\resizebox{1\columnwidth}{!}{%
  \includegraphics{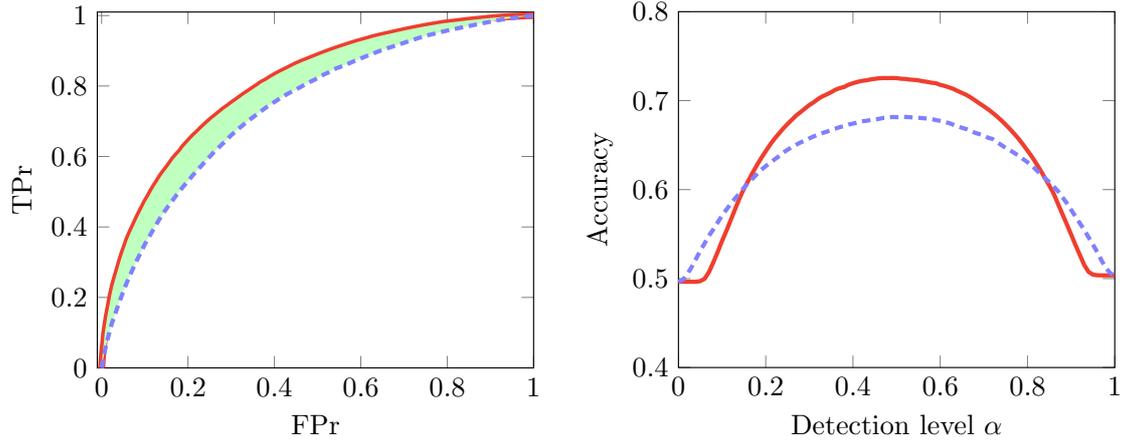}
}
\caption{Mean ROC curves for both models for the Dow Jones daily sampling (left) and the accuracy as a function of the detection level (right). The full line illustrates the memoryless model and the dashed line the historical one. These curves were obtained with a ten-fold cross-validation scheme. The shaded area illustrates the difference between AUC's.}
\label{fig:DJ}
\end{figure}

For the memoryless model, the AUC is equal to ($0.797\pm0.038$) and the mean maximum accuracy is equal to $73\%$.
For the historical model the AUC is equal to ($0.740\pm0.049$) and the mean maximum accuracy is equal to $68\%$. The difference between both AUC's is illustrated by the shaded area in the Fig-\ref{fig:DJ}. The predictive power is affected by the finite size estimation and the large number of parameters to be estimated (especially in the historical model).

To know if the timescale affects the predictive power, we performed the same analysis on a smaller timescale (3 orders of magnitude smaller).

\begin{figure}
\begin{center}
\resizebox{1\columnwidth}{!}{%
  \includegraphics{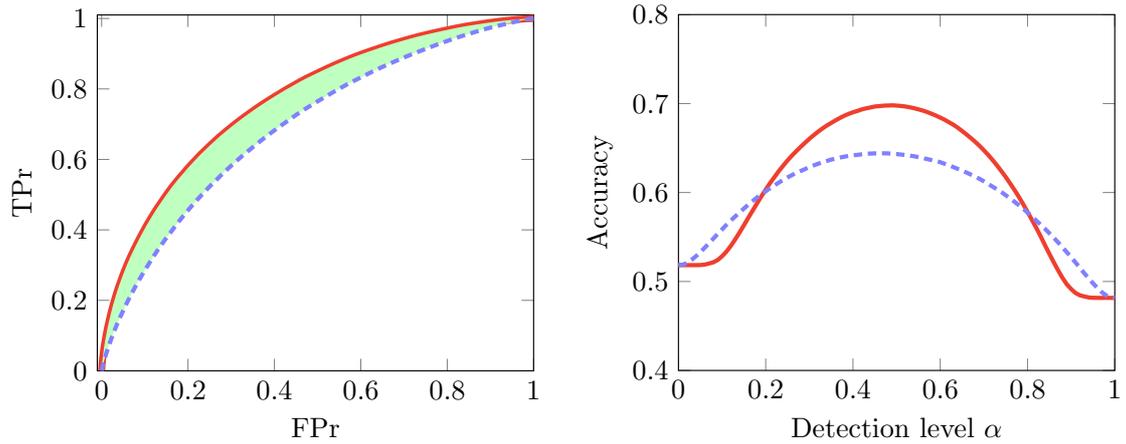}
}
\caption{Mean ROC curves for the Dow Jones minute sampling (left) and the accuracy as a function of the detection level (right). The full line illustrates the memoryless model and the dashed line the historical one. These curves were obtained with a ten-fold cross-validation scheme. The shaded area illustrates the difference between AUC's.}
\label{fig:DJmin}
\end{center}
\end{figure}

For the memoryless model, the AUC is equal to ($0.763\pm0.029$) and the mean maximum accuracy is equal to $70\%$.
For the historical model the AUC is equal to ($0.695\pm0.037$) and the mean maximum accuracy is equal to $64\%$. The difference between both AUC's is illustrated by the shaded area in the Fig-\ref{fig:DJmin}.

These values are slightly lower than in the daily sampling analysis, the relative difference between accuracy of both timescales is equal to $4 \%$. Moreover, the independent instantaneous model has an accuracy equal to $58\%$ significantly larger than for daily sampling results ($51\%$). These results are consistent with the observed lower correlation between returns at lower timescale (Epps effect) \cite{Epps}. The historical model is the least efficient. We conclude that the most significant part of the prediction model is the one capturing instantaneous co-movements.

Interestingly, the results are slightly improved if the pairwise influences $J_{ij}$ are set to their mean value (homogeneous influences) and if individual biases $h_{i}$ are set to zero. However the improvement is slight, the relative difference with the heterogeneous case is about $2\%$. For the Dow Jones at minute sampling, the resulting accuracy is equal to $71\%$, the AUC is equal to $(0.786 \pm 0.026)$. For the Dow Jones at daily sampling, the accuracy is equal to $73\%$, the AUC is equal to $(0.810 \pm 0.030)$. Given the relatively small width of the time-window, the reconstruction errors on these parameters induces biased results. Indeed, it was shown that the market structure is well described by a heterogeneous pairwise maxent model and not by a homogenous one \cite{moi1}.

\subsection{Dependencies on number of units, sample length and distance}

\begin{figure}
\begin{center}
\resizebox{0.75\columnwidth}{!}{%
  \includegraphics{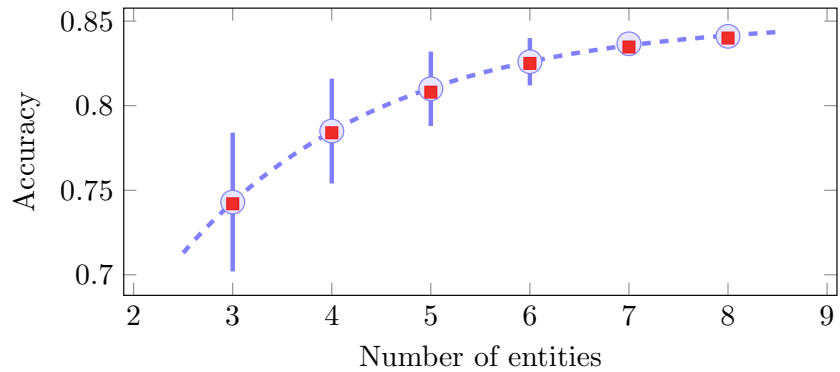}
}
\caption{Accuracy as a function of number of indices. Blue dots illustrate the accuracy of the instantaneous model and red squares the accuracy of the historical model. The dashed line is an exponential fit.}
\label{fig:AccN}
\end{center}
\end{figure}

The collective dynamics seems to be important for predicting flips. Adding more indices may improve the accuracy of the flipping detection. To study the dependency on system size, we let only $k$ indices visible among the $N=8$ European indices and we perform flipping prediction on the reduced system. For each value of $k$, we perform prediction on $N!/k!(N-k)!$ possible choices of indices set. Results are illustrated in Fig-\ref{fig:AccN}.


\begin{figure}
\begin{center}
\resizebox{0.75\columnwidth}{!}{%
  \includegraphics{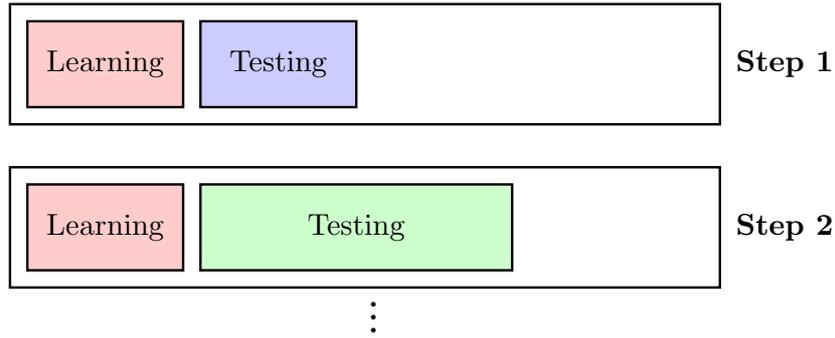}
}
\caption{Schematic description of the method to check accuracy dependence on the length of the testing block. We divide the sample in blocks. Lagrange parameters are inferred on the learning block. We use these parameters and  empirical data of the testing block to perform flipping prediction.}
\label{fig:learning1}
\end{center}
\end{figure}

The accuracy may also depend on the length of the testing sample. To check this feature, we infer Lagrange parameters with the rPML method on a learning block and we perform prediction on a testing block of increasing length. This method is illustrated in Fig-\ref{fig:learning1}.


\begin{figure}
\begin{center}
\resizebox{0.75\columnwidth}{!}{%
  \includegraphics{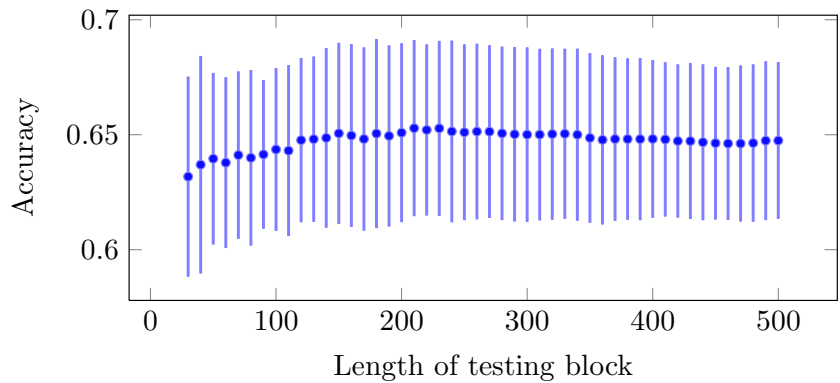}
}
\caption{Accuracy as a function of length of the testing block. Error bars represent the standard deviation on 8 different testing blocks. Parameters are inferred on a learning block of 500 samples and accuracy is measured on 8 different testing blocks, each of length increasing from 30 to 500 points (2 trading years) by increment of 10 samples.}
\label{fig:AccLength}
\end{center}
\end{figure}

\begin{figure}
\begin{center}
\resizebox{0.75\columnwidth}{!}{%
  \includegraphics{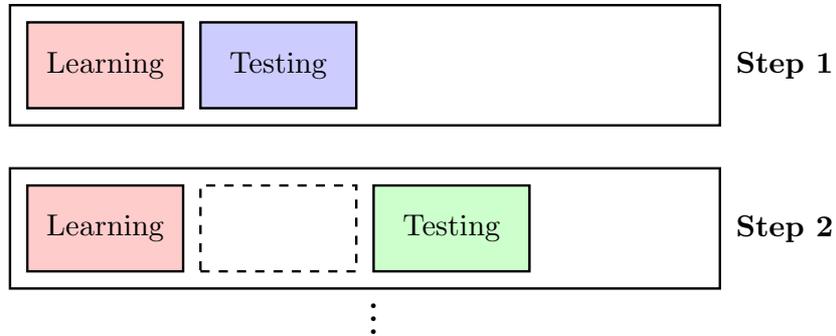}
}
\caption{Schematic description of the method to study the dependence on the distance between learning and testing blocks. We divide the sample in blocks. Lagrange parameters are inferred on the learning block. We use these parameters and  empirical data of the testing block the perform flipping prediction. Length of testing blocks is fixed.}
\label{fig:learning2}
\end{center}
\end{figure}

\begin{figure}
\begin{center}
\resizebox{0.75\columnwidth}{!}{%
  \includegraphics{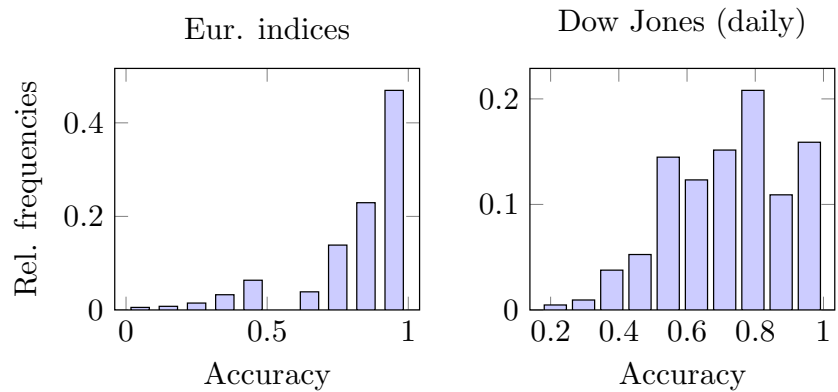}
}
\caption{Accuracy as a function of the distance between the learning and testing blocks for the Dow Jones index (1982-2000 period). Parameters inference is done on 1000 first points (1982-1985) and the accuracy is evaluated on 89 blocks of 40 points. The full line illustrates the instantaneous model and the dashed line the historical model.}
\label{fig:AccBlock}
\end{center}
\end{figure}

The accuracy seems to remain constant as the size of the testing block increases as illustrated in Fig-\ref{fig:AccLength}.
If the series was stationary, Lagrange parameters should be the same for the whole sample and we expect that the accuracy to be constant. For a non-stationary time series, Lagrange parameters may vary through time and so the accuracy. However if significant deviations from their mean values only occur on small time-windows, the accuracy appears constant when computed on large time-windows.
To study this feature, we test the dependency of the accuracy on the distance between learning and testing blocks. Instead of taking larger and larger testing blocks, we consider testing blocks of fixed length but farther and farther from the learning block. This procedure allows to compare accuracy on these different time-windows of fixed length. The schematic representation of this method is illustrated in Fig-\ref{fig:learning2} and results in Fig-\ref{fig:AccBlock}.

Returns exhibit volatility clustering, so we expect the accuracy will differ from its mean value only on small time-windows and we should observe a nearly constant value on a large time-window for fixed Lagrange parameters. In Fig-\ref{fig:AccBlock}, we observe that accuracy reaches its maximum value in the testing block embedding Black Monday (October 19, 1987). This feature is consistent with \cite{PreisQuant} where the largest mean correlation coefficient corresponds to the largest normalized return. A larger accuracy results from larger correlations during the crash. The difference between the maximum ($0.82$) and the minimum ($0.55$) accuracy is larger than the expected statistical error $40^{-1/2}\simeq0.16$, the increase of accuracy during crises is thus a genuine feature. A genuine temporal evolution of the accuracy is expected since the financial network can be depicted as a dynamical network with failures (crashes involving larger co-movements) and recoveries \cite{Majd}.

\begin{figure}
\begin{center}
\resizebox{0.75\columnwidth}{!}{%
  \includegraphics{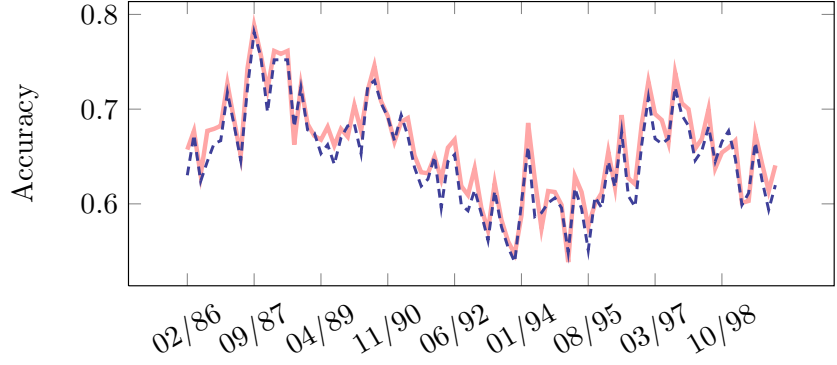}
}
\caption{Distribution of accuracy (averaged over the $N$ entities) over a time-window of 1000 trading days width for the European indices (left) and for the Dow Jones at daily sampling (right).}
\label{fig:AccFreq}
\end{center}
\end{figure}

Last, we note that over a time-window of 1000 trading days width, the averaged accuracy per trading day is rarely equal to zero as illustrated in Fig-\ref{fig:AccFreq}. For the European indices set, the averaged accuracy is equal to zero only for 6 trading days (31/08/2007, 18/10/2007, 22/04/2009, 14/04/2011, 06/02/2010, 23/02/2012). The first two occurrences happened just before the subprimes crisis, the third occurrence during the 2009 market rebound, the fourth at the end of the rebound following the Fukushima accident, the fifth and sixth happened during the recovery after the debt crisis (high risk periods). There is no obvious periodicity in the time series of accuracy (no fundamental frequency in the Fourier series). One can expect that Friday could be a day where accuracy decreases due to the expiration of securities but it is not observed in this analysis.

Another possibility is the one given in \cite{Livan11}: few driving forces can lead to a rich structure even in the bulk of the spectrum of the correlation matrix which is therefore not only due to noise. Such factors and clusters can also be thought as correlated structures appearing in the vicinity of the critical state of a pairwise maxent model. Global correlations (correlation length of the order of the network size) together with fluctuating clusters can coexist near the order-disorder boundary.

\section{Simultaneous trend reversals}
\begin{figure}
\begin{center}
\resizebox{0.9\columnwidth}{!}{%
  \includegraphics{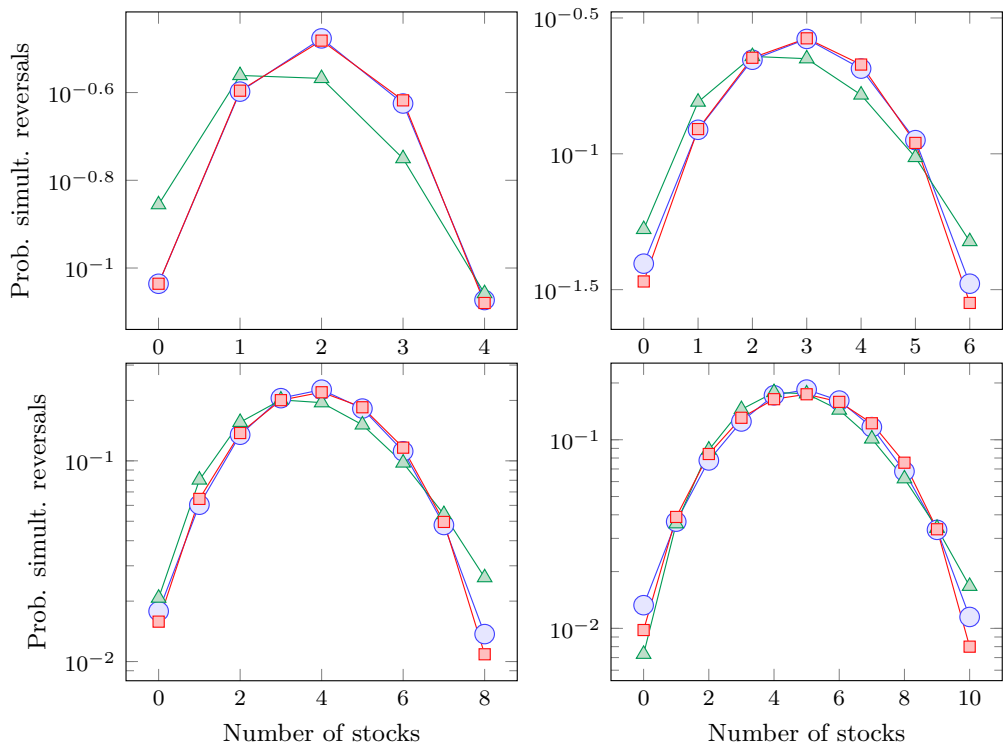}
}
\caption{The distributions of simultaneous trend reversals. The empirical distribution is illustrated by dots, the pairwise distribution by squares and independent Poissonian model by triangles. The distribution is computed over 20 randomly chosen sets (for $N=4, 6, 8, 10$ stocks from top left to bottom right) of the Dow Jones at minute sampling.}
\label{fig:simDist}
\end{center}
\end{figure}

We also inquire if the pairwise autologistic model is able to estimate the distribution of simultaneous trend reversals. The occurrence of a trend reversal is expressed by a binary variable $x_{i,t}=\mathbf{1}_{[s_{i,t+1}=-s_{i,t}]}$. Using the maximum entropy principle, we get the following pairwise maxent model

\begin{equation}
p_{2}(x_{1,t};\cdots;x_{N,t})=\mathcal{Z}^{-1}\exp\left(\sum_{i, j=1}^{N}W_{ij}x_{i,t}x_{j,t}\right)\label{Lagrange2}
\end{equation}
where the matrix $\textbf{W}$ has a non null diagonal and can be estimated by the method detailed in \cite{Dick}. We also fit an independent trend reversal model (a Poisson distribution, using the maximum likelihood estimator). We compare the empirical, pairwise and independent distributions on 10 randomly chosen groups for different sizes (up to $N=12$ where direct sampling gives a good estimate of the distribution). Results are illustrated in Fig-\ref{fig:simDist}. The most frequent event is a reversal of approximatively half of the number of considered stocks.

\begin{figure}
\begin{center}
\resizebox{0.95\columnwidth}{!}{%
  \includegraphics{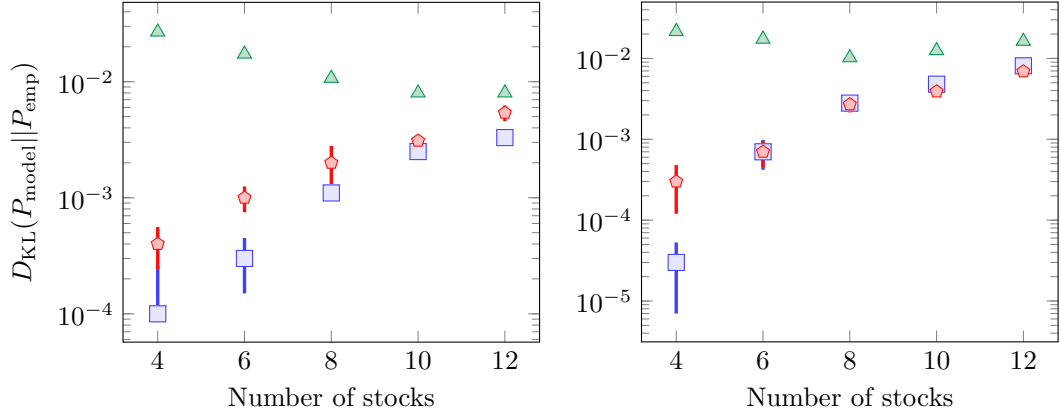}
}
\caption{The average Kullback-Leibler divergence between the empirical distribution of simultaneous reversals and the pairwise (squares), independent (triangles) and dichotomized Gaussian (pentagons) models. The divergence is computed over 10 randomly chosen stock sets of the Dow Jones at daily sampling (left) and at minute sampling (right). Error bars represent the standard deviation over 10 randomly chosen stock sets.}
\label{fig:DKL}
\end{center}
\end{figure}

We computed the mean Kullback-Leibler divergence between the empirical distribution and the pairwise, independent and dichotomized Gaussian models. The results are illustrated in Fig-\ref{fig:DKL}. The pairwise model is the closest to the empirical distribution.

The dichotomized Gaussian (DG) model \cite{Amari,Macke09} is a threshold multivariate Gaussian model with mean and covariance matrix inferred to match the empirical first and second moments of the binary time series. It is an attractive alternative to the pairwise maxent model because the parameters are easier to infer and it can be used to characterize higher-order interactions \cite{Yu11}. As illustrated in Fig-\ref{fig:DKL}, its accuracy of simultaneous reversals prediction is similar to the one of the pairwise maxent model. Therefore, there is no reason to rule out the pairwise maxent. This result is consistent with the multi-information criterion which returns that pairwise statistical dependencies represent $95\%$ of statistical dependencies \cite{moi1}.


\section{Conclusion}
Our results suggest that trend reversals can be predicted using instantaneous collective states of other market places. This finding also reveals the strength of the collective dynamics underlying the flipping process since the individual instantaneous model is not able to make better than random predictions excepted at higher sampling frequency. Another advantage is that this pairwise maxent model satisfies \emph{all} the pairwise correlations simultaneously which can prevent the overcounting of dependencies using only the pairwise correlation when more than two entities are involved.
Including memory in this model does not improve the accuracy of prediction. This is a not very surprising result since the pairwise lagged cross-correlations are close to zero. Moreover, the sign of returns is poorly forecast ($53\%$ of accuracy) when we use only returns past information. This result is inline with the efficient market hypothesis and a profit cannot be made using this model. However, the sample length is too small to estimate so many parameters.  The history may be important in more evolved model including a temporal filtering on the basis of a good approximation of the market dynamics (by analogy to the treatment of time series, especially in the neuroscience field) or modelling with exogenous economic variables. An interesting interpretation of the fine structure of the spectrum of the correlation matrix \cite{Livan11} is that such models allow global correlations (with characteristic length of the order of the network size) and fluctuating clusters coexist in the vicinity of the critical state \cite{Fischer}. This may account for the global collective mode, corresponding to the largest eigenvalue of the correlation matrix, and to the structure of the spectrum bulk which is not only due to noise but also account for clustering properties \cite{Livan11}.

It is interesting that such a minimal model returns an accuracy almost as good than the accuracy of pairwise autologistic models even if the market dynamics is undoubtedly much more complex than the model; this finding highlights the significant contribution of collective modes in trend prediction since individual biases are non relevant for the prediction excepted at higher sampling frequencies. Furthermore, it suggests that methods derived in neuroscience could be also applied in finance even if the couplings seem to scale differently \cite{moi2,Schneidman}.

\section*{Acknowledgments}
I would like to thank D. Veredas, B. De Rock, P. Emplit, A. Smerieri and S. Massar for their helpful comments and discussions. This work was undertaken with financial support from the Solvay Brussels School of Economics and Management.

\appendix
\section{Cleaning the data}
In this work, we consider instantaneous information (within the defined time bin). The time series should therefore be synchronous. The stock exchange closing days, pre-market and after hours trading exchanges are removed. If a time bin is missing for a particular asset, the same time bin should be deleted from the database. The later case is marginal since we consider indices and highly capitalized companies.
\section{Regularized pseudo-maximum likelihood}

The rPML method is a powerful method for the estimation of Lagrange parameters of pairwise maximum entropy model when common maximum likelihood is untractable \cite{Aurell}. This method can be thought as an autologistic regression in order to predict binary outcomes. The main idea is to factorize the distribution and to consider only conditional probabilities. For a N-dimensional sample of length $T$, the objective function to maximize is

\begin{equation}\label{PML}
  \mathrm{PL}(\boldsymbol\theta)=\frac{1}{T}\sum_{t=1}^{T}\sum_{i=1}^{N}
  \log P(s_{i,t}|\mathbf{s}_{-i,t};\, \boldsymbol\theta)
\end{equation}

where conditional probabilities of the instantaneous model are

\begin{equation}
p(s_{i,t}|\mathbf{s}_{-i,t};\, \boldsymbol\theta)=\frac{1}{2}
\left[1+s_{i,t}\tanh\left(\sum_{ j\neq i}J_{ij}s_{j,t}+h_{i}\right)\right]
\end{equation}


and

\begin{equation}
p(s_{i,t}|\mathcal{H}_{t}^{T};\, \boldsymbol\theta)=\frac{1}{2}
\left[1+s_{i,t}\tanh\left(\sum_{ j\neq i}J_{ij}s_{j,t}+h_{i}
+\sum_{\tau=1}^{T}\sum_{j}K_{ij}^{\tau}s_{j,t-\tau}\right)\right]
\end{equation}


for the historical model.

A regularization term is added to the PL function to prevent overfitting which is a negative multiple of the $l_{2}$-norm of parameters to be estimated, for instance. The regularized PL (rPL) objective function is thus $\mathrm{PL}(\boldsymbol\theta)-\lambda\, \|\boldsymbol\theta\|_{2}^{2}$ with $\lambda>0$. If the network is believed to be sparse, a $l_{1}$ regularization term should be used \cite{Aurell} (small values of the parameters are projected on zero).

\section{Noise and comparison to artificial networks}
The estimation of the Lagrange parameters may introduce a bias in orientation prediction. Particularly because of noise due to finite size estimation and limitation of inference methods based on approximation scheme. Moreover, their values depend on the considered sample since they  are such that the first and second moments should match empirical ones.

\begin{figure}
\begin{center}
\resizebox{0.75\columnwidth}{!}{%
  \includegraphics{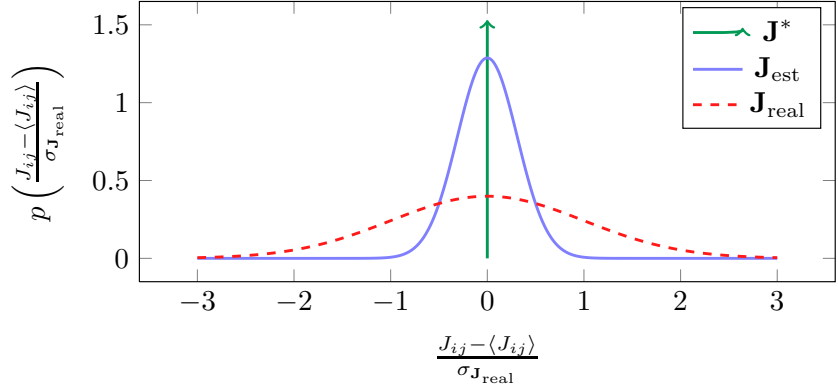}
}
\caption{Schematic representation of noise level estimation in parameters inference. Artificial data are generated with homogeneous influences $\textbf{J}^{*}$ (probability density function illustrated by the green Dirac delta). Then we perform parameters estimation using these artificial data. Ideally the pdf of the estimated parameter $\textbf{J}_{\mathrm{est}}$ should be close to the pdf of $\textbf{J}^{*}$. Last, we compare the distribution of $\textbf{J}_{\mathrm{est}}$ to the variance of parameters resulting from real data $\textbf{J}_{\mathrm{real}}$ using their variance.}
\label{fig:SchNoise}
\end{center}
\end{figure}

To quantify the bias, we estimate the noisy part of the standard deviation of the recovered $\textbf{J}$ matrix. We simulate binary time series (same sample length as the true data) with the maximum entropy conditional probability $p(s_{i,t}=-s_{i,t-1}|\mathbf{s}_{-i,t})$, known as the Glauber dynamics \cite{BinderMC}. A product is randomly chosen, a flipping attempt is accepted if the flipping probabilities $2^{-1}[1-s_{i}\tanh(\sum_{j}J_{ij}^{*}s_{j})]$ is larger than a randomly uniform number on the interval $[0,1]$. A configuration is recorded each Monte Carlo step (MCS). A MCS corresponds to $5N$ flipping attempts. In this data generation, the artificial $\textbf{J}^{*}$ matrix was taken homogeneous with all entries equal to the empirical mean of mutual influences. Then we estimate the influence matrix with the rPML method. Ideally, the standard deviation $\sigma_{\mathrm{noise}}$ of the estimated artificial influences should be much smaller than the one of real influences $\sigma_{J}$. The method is schematically illustrated in fig-\ref{fig:SchNoise} and results are reported in Table-\ref{tab:noise}. Depending on the sample length, the noise seems to be significant but not the dominant part of the estimation except for large system size.

\begin{table}[!ht]
\caption{Quantification of the noisy part of the standard deviation of the inferred mutual influences.}
\label{tab:noise}
\begin{center}
\begin{tabular}{lll}
\hline
Index/set    & sample length ($T$) & $\sigma_{\texttt{noise}}/\sigma_{J}$\\
\hline
Eur. indices &  $2.5\times 10^{3}$ & 0.37                                \\
DJ(daily)    &  $2.5\times 10^{3}$ & 0.31                                \\
DJ(min)      &  $3.0\times 10^{4}$ & 0.24                                \\
\hline
\end{tabular}
\end{center}
\end{table}


We can also generate data with the estimated $\textbf{J}$ matrix from the data, infer the artificial $\textbf{J}^{*}$ matrix and compare $\textbf{J}^{*}$ to $\textbf{J}$. The reconstruction is satisfying if estimated Lagrange parameters $J_{ij}^{*}$ are close to their true values $J_{ij}$. To quantify deviation from the real network (defined by $\textbf{J}$), we use the reconstruction error $\Delta=\sqrt{N}\langle(J_{ij}^{*}-J_{ij})^{2}\rangle^{1/2}$ which represents the ratio between the root mean square error $\langle(J_{ij}^{*}-J_{ij})^{2}\rangle^{1/2}$ and a canonical standard deviation $1/\sqrt{N}$ \cite{Aurell}. This definition of the reconstruction error is believed to be consistent with financial networks \cite{moi2}. Results are reported in Table-\ref{tab:rplm}. These results are consistent with those of \cite{Aurell} where the magnitude order of the reconstruction error is $10^{-2}$ for a complete network of size $N=64$ with $J_{ij}$ drawn from a Gaussian distribution $\mathcal{N}(0,N^{-1})$.

\begin{table}[!ht]
\caption{Quantification of the reconstruction error $\Delta$ with the regularized pseudo-likelihood. Artificial data are generated with the Glauber dynamics using $\textbf{J}$ inferred from real data as true influences matrix (a configuration was recorded each $5N$ flipping attempts).}
\label{tab:rplm}
\begin{center}
\begin{tabular}{lll}
\hline
Index/set             & sample length ($T$) & $\Delta$\\
\hline
Eur. indices       &  $2.5\times 10^{3}$ & 0.100                           \\
DJ(daily)          &  $2.5\times 10^{3}$ & 0.158                           \\
DJ(min)            &  $3.0\times 10^{4}$ & 0.035                           \\
DJ(min)            &  $1.0\times 10^{6}$ & 0.026                           \\
\hline
\end{tabular}
\end{center}
\end{table}


A useful benchmark to assess exactness of this autologistic model may be the predictive power computed from artificial data. We compute the mean accuracy and mean AUC for artificial data truly generated by a pairwise autologistic process and we compare them to the results obtained from financial data. These values are reported in Table-\ref{tab:Art}.

\begin{table}[!ht]
\caption{Comparison of artificial accuracy and AUC to real ones. The artificial values are computed from data generated with a pairwise maximum entropy model (autologistic) and the real ones from financial data. Artificial samples are of the same length than the corresponding real samples.}
\label{tab:Art}
\begin{center}
\begin{tabular}{lcccc}
\hline
Index/set            & Accuracy art. ($\%$)& Accuracy ($\%$)& AUC art.& AUC   \\
\hline
Eur. indices         & 87                  &  83            & 0.911   & 0.914 \\
DJ(daily)            & 75                  &  73            & 0.806   & 0.797 \\
DJ(min)              & 71                  &  70            & 0.769   & 0.763 \\

\hline
\end{tabular}
\end{center}
\end{table}


In general, the predictive power is slightly larger for artificial data. The relative difference between artificial and real data lies between $ 1\%$ and $5\%$. This benchmark reveals that sign of returns can be predicted with similar accuracy than finite size time series truly generated by a pairwise instantaneous process. The artificial accuracy and AUC represent the maximum expected values that the model can return due to the finite size effects.



\end{document}